\documentclass[useAMS,usenatbib]{mn2e}

\usepackage{graphicx}

\title[Filament Shape and Coronal Magnetic Field Structure]
{Filament Shape Versus Coronal Potential Magnetic Field Structure}
\author[B. Filippov]{B. Filippov \thanks{E-mail:
bfilip@izmiran.ru}  \\ Pushkov Institute of Terrestrial Magnetism,
Ionosphere and Radio Wave Propagation of the Russian Academy of
Sciences (IZMIRAN), \\ Troitsk, Moscow 142190, Russia}

\begin{document}

\date{Accepted 0000 December 15. Received 0000 December 14; in original form 0000 October 11}

\pagerange{\pageref{firstpage}--\pageref{lastpage}} \pubyear{2002}

\maketitle

\label{firstpage}

\begin{abstract}
Solar filament shape in projection on disc depends on the
structure of the coronal magnetic field. We calculate the position
of polarity inversion lines (PILs) of coronal potential magnetic
field at different heights above the photosphere, which compose
the magnetic neutral surface, and compare with them  the
distribution of the filament material in H$\alpha$ chromospheric
images. We found that the most of the filament material is
enclosed between two polarity inversion lines (PILs), one at a
lower height close to the chromosphere and one at a higher level,
which can be considered as a height of the filament spine.
Observations of the same filament on the limb by the {\it STEREO}
spacecraft confirm that the height of the spine is really very
close to the value obtained from the PIL and filament border
matching. Such matching can be used for filament height
estimations in on-disk observations. Filament barbs are housed
within protruding sections of the low-level PIL. On the base of
simple model, we show that the similarity of the neutral surfaces
in potential and non-potential fields with the same
sub-photospheric sources is the reason for the found tendency for
the filament material to gather near the potential-field neutral
surface.
\end{abstract}


\section{Introduction}

Solar prominences are 'clouds' of cooler and denser plasma in the
much hotter and rarefied solar corona. Most of them look like long
dark 'filaments' in strong spectral lines due to absorption of the
background radiation when they are projected on the solar disk.
Quiescent prominences (filaments) frequently take the form of thin
ribbons or curtains. The typical thickness of the ribbons (5-6 Mm)
is much smaller than both their height (30-35 Mm) and their length
along the solar surface ($> 200$ Mm) \citep{b11, b50, b44, b33,
b39}. Filament locations on the Sun reflect their magnetic origin.
Comparisons with maps of the photospheric magnetic field show that
filaments are always located above lines separating opposite
polarities of the radial field, polarity inversion lines (PIL)
sometimes called neutral lines \citep{b6, b24, b48, b37}.

A region surrounding a PIL forms an inversion zone that
corresponds to the filament channel. A simple potential
extrapolation of the photospheric field yields an arcade of loops
in this inversion zone. The magnetic field-lines directly above
the inversion line should be horizontal and predominantly
transverse to this line. In fact, the field in the filament
channel is much more complex. Zeeman \citep{b57, b45, b26, b49,
b27}  and Hanle \citep{b30, b31, b7, b8}  effect measurements of
the magnetic fields in prominences have shown that, as a rule, the
field in the channel is horizontal, but directed nearly along the
filament axis and PIL. The angle to the axis is 25$^\circ$, on
average. Moreover, the component transverse to the axis is
predominantly opposite to that calculated in the potential
approximation.

To support stably heavy cool plasma against gravity, a magnetic
field lines should be locally horizontal and curved upward, i.e.,
should contain dips in the field line shapes \citep{b28}. In
particular, dips are present in a potential field with a
quadrupolar structure of underlying photospheric magnetic sources.
However, many observational details indicate that the magnetic
fields in filament channels are not potential and contain
significant electric currents. Two non-potential magnetic
configurations are usually considered as possible 'magnetic
skeleton' of filaments. Magnetic flux ropes with nearly force-free
field that lie horizontally above the PIL have dips in lower parts
of helical field lines \citep{b29, b41, b52, b43, b46, b2, b9,
b22}. Sheared arcades are also proposed as magnetic structures of
the filament channels \citep{b1, b12, b5}.

Typically 3 major structural components can be recognized in a
filament, namely a spine, barbs, and two extreme ends. The spine
is a nearly horizontal line along the top of the filament. The
barbs or intermediate 'legs' of the filament protrude from its
main body on both sides. The barbs look to connect the spine to
the chromosphere below, when the filament is observed closer to
the limb.

When observed from above, the filament barbs are seen to protrude
at an acute angle either clockwise or counterclockwise with
respect to the axis of the filament \citep{b36}. Depending on the
deviation of the barbs, filaments were classified as either
right-bearing or left-bearing. In exactly the same way, filaments
can be classified as dextral or sinistral depending on the
direction of the axial component of the field to the right or to
the left, when viewed from the major photospheric positive
polarity \citep{b36}. This handedness property was called
'chirality' \citep{b34}. Martin and collaborators found that
dextral filaments have always right-bearing barbs, while sinistral
filaments have left-bearing barbs.

Many authors studied the relationship between the position of the
filament barbs and the underlying photospheric magnetic field.
Some results are contradictory possibly because the real height of
the barb endpoints is uncertain in on-disk observations.
\citet{b40} found that most of filament feet are clustered in
junctions of three or more adjacent supergranular cells, where the
concentration of vertically oriented field is the largest.
\citet{b36} and \citet{b51} believed that the ends of the barbs
are connected to weak magnetic fields in between the network
elements. \citet{b35} found that the barbs are anchored in
parasitic (minority) polarity elements. \citet{b54, b55},
\citet{b10}, and \citet{b32}  showed that the barbs are ended very
close to small-scale PILs between majority and minority polarities
on the side of the filament.

Since the vertical size of barbs is much larger than the
gravitational scale height of the prominence plasma, the barbs
cannot be static structures within inclined flux tubes. The
prominence plasma must somehow resist to falling down to the
chromosphere or the prominence must be a non-equilibrium dynamic
formation. Therefore, the structure of barbs and the orientation
of the magnetic field in them are key issues. Aulanier and
collaborators \citep{b2, b3, b4}  developed a model of
three-dimensional magnetic configuration of filaments based on the
idea of magnetic dips, which successfully reproduced the general
shape of filaments. They used either linear force-free fields or
linear magnetohydrostatic fields. Plasma of the main body of the
filament is supported at local dips in the bottom part of the
helical windings of a coronal flux rope. Barbs are formed in small
dips above minor photospheric polarity inversion lines around
magnetic elements whose polarity is opposite to the dominant
polarity of surrounding magnetic fields. \citet{b51}  constructed
a three-dimensional magnetic model of a filament using a nonlinear
force-free configuration. The initial flux rope was subjected to
magnetofrictional relaxation. The dips in the helical field lines
are formed under the influence of neighboring network elements.
The model reproduces the observed filament barb.

A photospheric PIL is the intersection of the surface $B_r = 0$,
which can be called a 'neutral surface', with the photosphere.
\citet{b58}  found that filaments not only follow photospheric
PILs but most of the filament material is concentrated close to
the neutral surface or surface passing through apices of arches of
potential magnetic field-lines. This property possibly appears due
to the fact that, while the neutral surface is not as a rule the
location of dips, it is a place where coronal currents can find
horizontal equilibrium. The axis of the flux rope containing a
filament should lie in the neutral surface and after the lost of
equilibrium during an eruption, the flux rope some time moves
along the neutral surface \citep{b20, b21, b14, b16}.

In this paper, we compare in more detail the shape of filaments
and the shape of neutral surfaces calculated in potential
approximation on the base of photospheric magnetograms. From this
comparison we are able to determine the height of filament spines
in on-disk observations. The results are verified by simultaneous
on-limb observations from a different viewpoint. Folds of the
neutral surface near the chromosphere revealed by projections of
low-altitude PILs are possible locations of filament barbs.

\section[]{Data and Method of Analysis}

We used full-disk H$\alpha$ filtergrams from the archive of the
Big Bear Solar Observatory for selecting rather wide filaments
with conspicuously protruding barbs when they were located not far
from the center of the solar disk or at least of the central
meridian. This limitation arises from necessity to have good
magnetographic data for a boundary condition in coronal magnetic
field calculations. Polar crown filaments, which often have
prominent barbs, are not likely for this purpose because
photospheric magnetic field measurements are too noisy in polar
regions.

\begin{figure*}
\includegraphics[width=167mm]{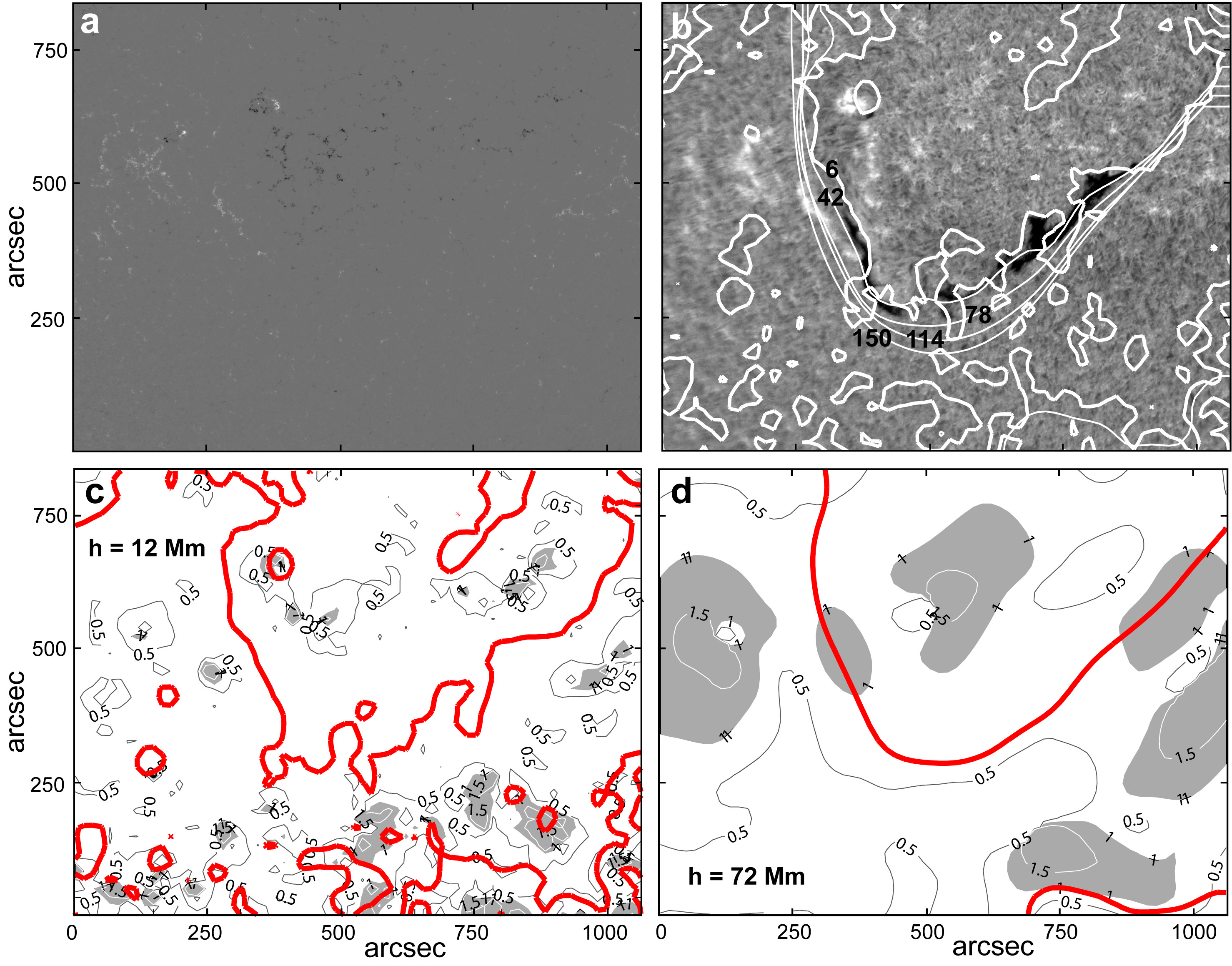}
\caption{(a) HMI magnetogram of the region taken around the
filament on 2012 February 23 at 20:21 UT. (Courtesy of the
NASA/{\it SDO} and the HMI science team). (b) H$\alpha$ filtergram
of the same region and at the same time with superposed PILs at a
height of 6 (thicker lines), 42, 78, 114, and 150 Mm (thinner
lines from top to bottom). (Courtesy of the Big Bear Solar
Observatory), (c) - (d) Distributions of the decay index and PILs
(thick red lines) at two heights. Shadowed areas show the regions
where $n > 1$.  }
\end{figure*}

Fig. 1a represents a fragment of the magnetogram taken by the
Heliospheric and Magnetic Imager (HMI; \citet{b47}  on board the
{\it Solar Dynamic Observatory} ({\it SDO}) on 2012 February 23 at
20:21 UT, which was used as the boundary condition for the
potential magnetic field calculations. We need the potential
magnetic field distribution in the corona at heights of
prominences, which are much less than a solar radius. Therefore,
we can use a restricted area of a photospheric magnetogram as the
boundary of the calculation domain and neglect its sphericity
considering it as a part of a flat surface \citep{b18, b19, b14} .
When we cut out a rectangular area around the filament under study
from the full disk magnetogram, we ignore the contribution of the
magnetic sources outside of it. Such simplification is reasonable
if the main sources of the field lie within the cut-out area. For
regions rather far from the center of the solar disk, we construct
the data array with the equal angular size of pixels and assume
the projection of the line-of-sight-field on the normal as the
radial component. Thus we obtain the boundary condition for the
calculation of the potential magnetic field as if the region were
located at the center of the disk.

We calculate $B_z$ maps ($z$-axis is vertical) at different
heights and obtain a set of PILs $B_z = 0$ as shown in Fig. 1b.
This set of curves represents the neutral surface nearby the
filament. More precisely, the set of curves shows the projection
of the neutral surface on the plane of the photosphere. Each curve
is the intersection of the surface $B_z = 0$ with a plane $z$ =
const. We choose the lowest level as the level of the middle
chromosphere ($\sim$ 5 Mm) in order to have a rather smooth PIL
without influence of small-scale fields. To take into account the
angle between the local vertical with the line-of-sight, each PIL
is displaced in a map along the $x$ and $y$ coordinates at
distances

\begin{equation}
\Delta x = h \tan \lambda_0 ,
\end{equation}
\begin{equation}
\Delta y = h \tan \varphi_0 ,
\end{equation}
where $h$ is the height of the PIL above the photosphere,
$\lambda_0$ is the longitude of the center of the area,
$\varphi_0$ is the latitude of it.

The set of curves we superpose on the H$\alpha$ filtergram of the
same region as was selected for the magnetic field calculations
(Fig. 1b). We choose a curve nearest to the filament spine. Most
of filament material is located between the lowest PIL and the PIL
that we believe correspond to the height of the filament spine.
For the filament shown in Fig. 1b this is the PIL at a height of
60 Mm.

To be sure that the filament can be in stable equilibrium at the
height that we obtain for the filament spine (60 Mm), we calculate
the distribution of the decay index \citep{b18, b19, b14}

\begin{equation}
n = - \frac{\partial \ln B_t}{\partial \ln h},
\end{equation}
of the horizontal potential magnetic field $B_t$ in this region at
different heights $h$ (Figs 1c-d) and find that the decay index is
less than unity at the PILs up to the height of 70 Mm. Therefore,
the flux rope containing the filament can be in stable equilibrium
if its axis, which usually corresponds to the filament spine, is
below this height.

\section[]{Filament Shapes and Polarity Inversion Lines }

\subsection[]{Filament on 2012 February 23}

\begin{figure*}
\includegraphics[width=167mm]{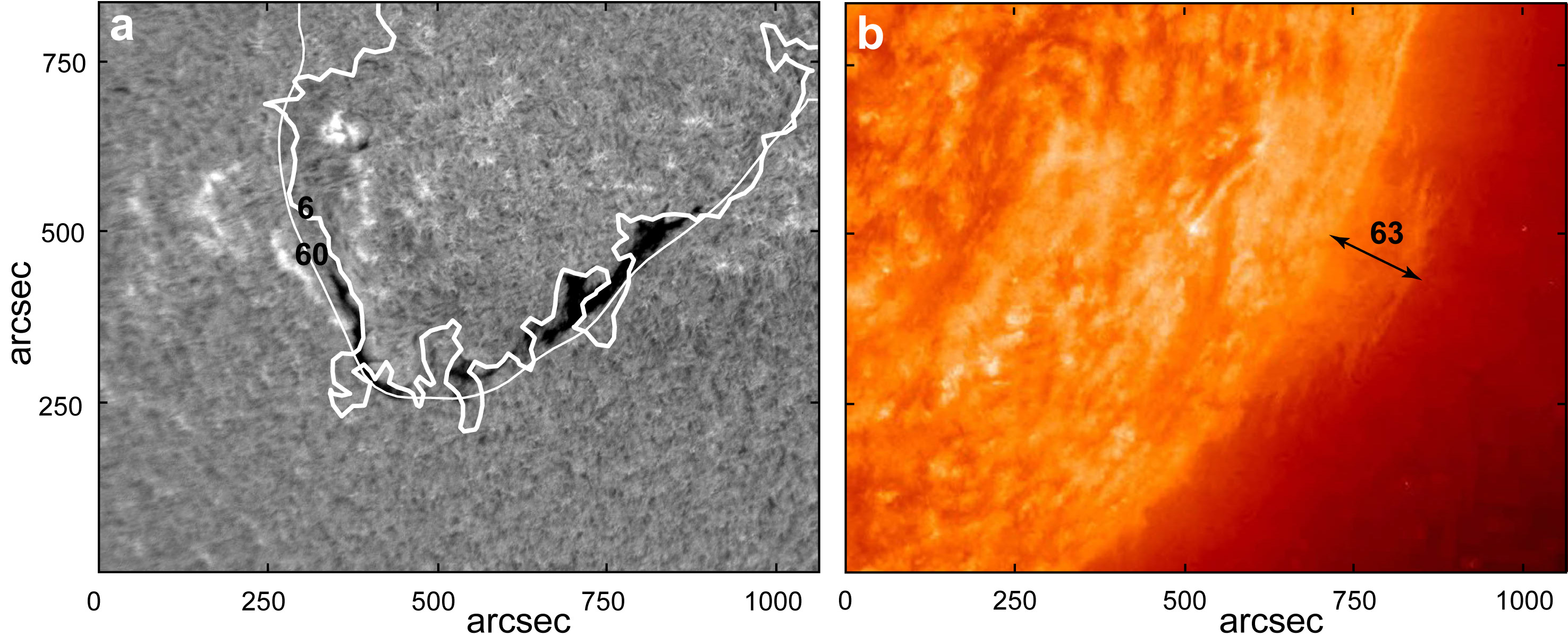}
\caption{(a) The same H$\alpha$  filtergram as in Fig. 1b with
superposed PILs related to the filament at heights of 6 (thicker
line) and 60 Mm (thinner line). (Courtesy of the Big Bear Solar
Observatory) (b) The same filament observed with {\it STEREO
B}/SECCHI EUVI 304-\AA \ on the limb on 2012 February 21 at 12:06
UT. (Courtesy of the {\it STEREO}/SECCHI Consortium.)}
\end{figure*}

Fig. 2a shows the same H$\alpha$ filtergram as in Fig. 1b with
superposed PILs at a height of 6 and 60 Mm. In all figures in this
section, we remove for clarity all PIL laying outside the
filaments and show only PILs closest to the filaments on both
sides (compare Fig. 2a and Fig. 1b). The PIL at the height of 60
Mm is nearest the filament spine. We can conclude that the height
of the spine is about 60 Mm. This is the case because the filament
was observed on the limb with the Sun Earth Connection Coronal and
Heliospheric Investigation (SECCHI) EUVI \citep{b56, b25}  on
board the {\it Solar Terrestrial Relations Observatory Behind}
({\it STEREO B}) on 2012 February 21 and with the identical
instrument on board {\it STEREO Ahead (A)} on 2012 February 24.
The separation angle with the Earth was 109$^\circ$ for {\it
STEREO A} and 117$^\circ$ for {\it STEREO B}. The height of the
highest point of the prominence spine (Fig. 2b) measured above the
limb is 63 Mm on both days.

The whole filament body in Fig. 2a is enclosed between the 6 and
60 Mm PILs. While the filament spine is aligned with the smooth
PIL at the height of 60 Mm, filament barbs are located between the
PILs. Ends of barbs touch protruding sections of the PIL at the
height of 6 Mm. This corresponds to their small heights above the
chromosphere. The low-altitude PIL outlines roughly the barbs.
However not all PIL protrusions are filled with filament barbs.
Since the filament is located at middle latitude in the southern
hemisphere, only barbs on the northern side of the filament are
visible.

\subsection[]{Filament on 2013 March 13}

\begin{figure*}
\includegraphics[width=167mm]{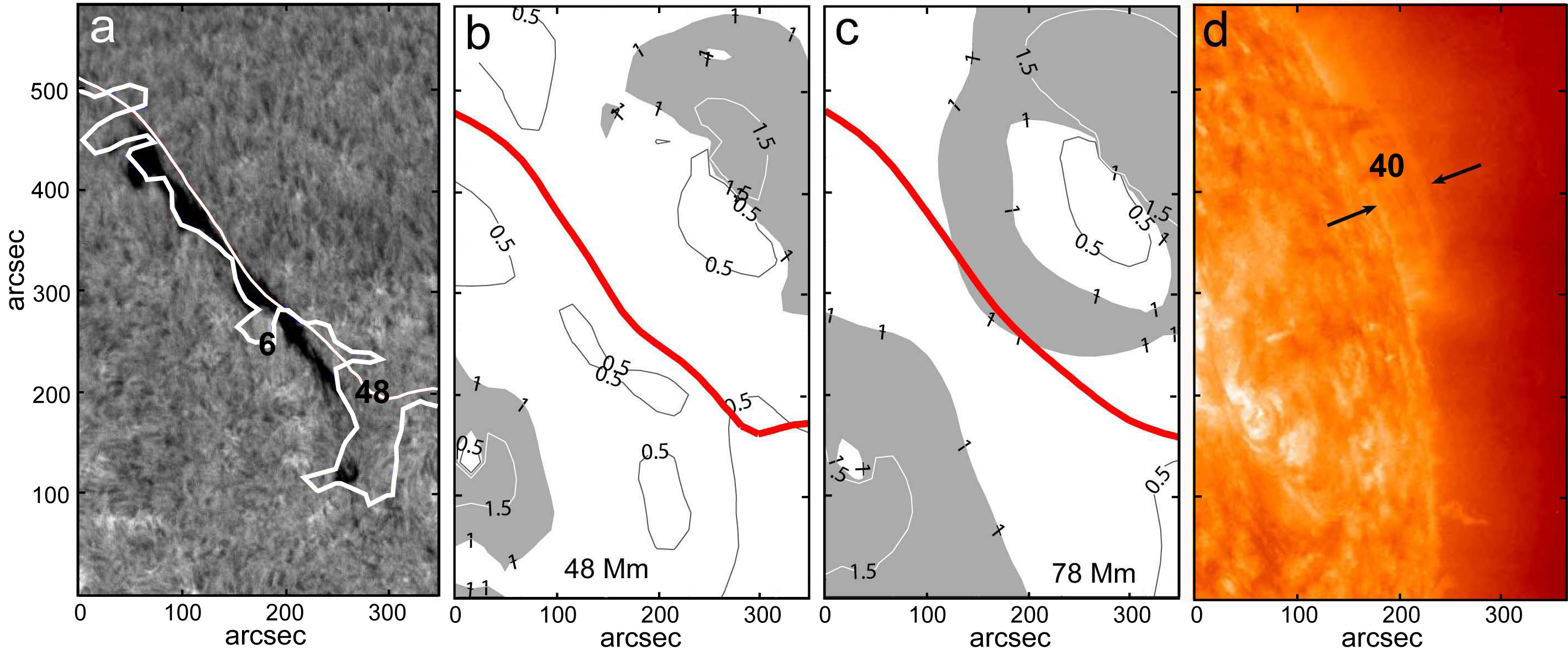}
\caption{(a) H$\alpha$ filtergram of the filament on 2013 March 13
at 16:14 UT with superposed PILs related to the filament at
heights of 6 (thicker line) and 48 Mm (thinner line). (Courtesy of
the Big Bear Solar Observatory). (b) - (c) Distributions of the
decay index and PILs (thick red lines) at different heights.
Shadowed areas show the regions where $n > 1$. (d) The same
filament observed with {\it STEREO B}/SECCHI EUVI 304-\AA \ on the
limb on 2013 March 09 at 11:46 UT. (Courtesy of the {\it
STEREO}/SECCHI Consortium.)}
\end{figure*}

A filament with the straight spine stretched from the north-east
to the south-west at middle latitudes in the northern hemisphere
was observed near the central meridian on 2013 March 13. The spine
fits best the PIL at a height of 48 Mm (Fig. 3a). Barbs on the
southern side of the filament fill the space between this PIL and
the PIL at a height of 6 Mm. The spine deviates from the 48 Mm PIL
at the southern end, possibly because the height of the spine
descends there. The height of the spine measured on the limb in
the {\it STEREO B}/SECCHI EUVI 304-\AA\  image on 2013 March 09 at
11:46 UT is 40 Mm (Fig. 3d). Distribution of the decay index
indicates that the filament is stable up to a height of 75 Mm
(Figs 3b-c).

\subsection[]{Filament on 2013 January 31 in the southern hemisphere}

\begin{figure*}
\includegraphics[width=167mm]{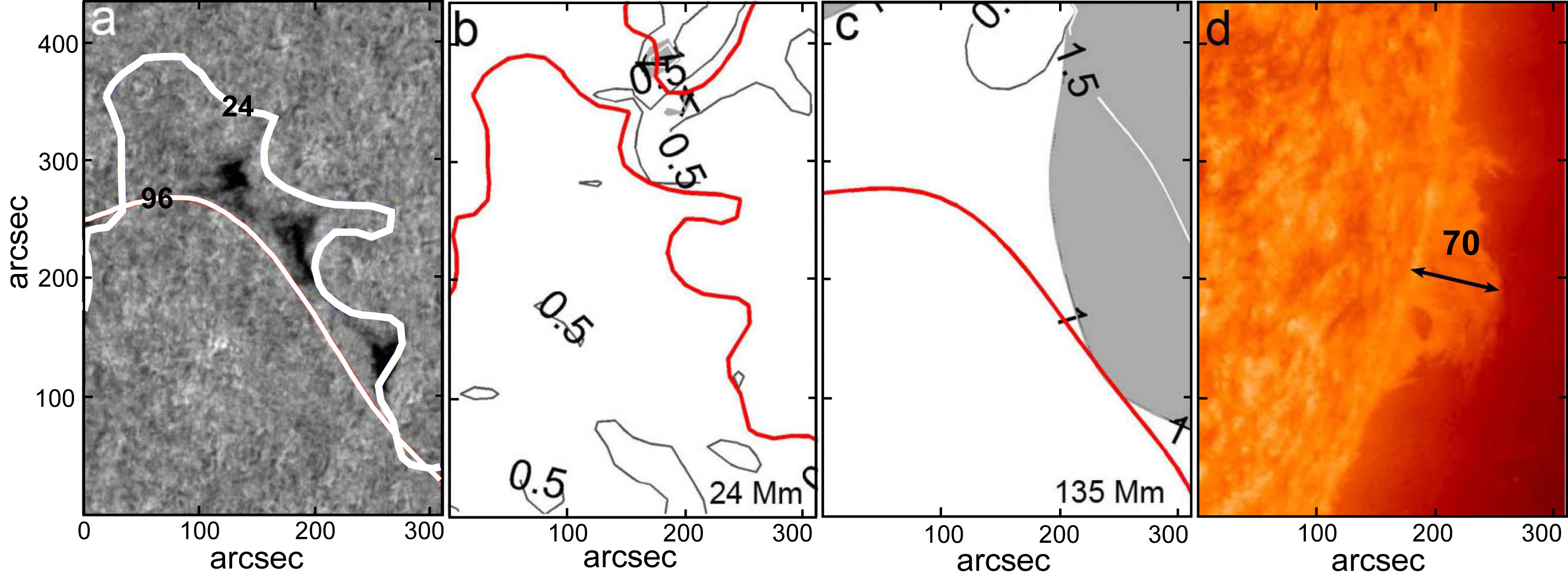}
\caption{H$\alpha$ filtergram of the filament on 2013 January 31
at 20:38 UT with superposed PILs related to the filament at
heights of 24 (thicker line) and 96 Mm (thinner line). (Courtesy
of the Big Bear Solar Observatory). (b) - (c) Distributions of the
decay index and PILs (thick red lines) at different heights. (d)
The same filament observed with {\it STEREO B}/SECCHI EUVI
304-\AA\ on the limb on
 2013 January 28 at 11:26 UT. (Courtesy of the {\it STEREO}/SECCHI
Consortium.)}
\end{figure*}

Two wide filaments were located close to the central meridian not
far from the equator in the southern and northern hemispheres on
2013 January 31. Both filaments have specific triangular barbs and
thin spines.

The southern filament has three prominent barbs (Fig. 4a), which
are enclosed between PILs at heights of 24 and 96 Mm. Lower than
24 Mm PILs deviates too far from the filament. The height of the
spine measured on the limb in the {\it STEREO B}/SECCHI EUVI
304-\AA\ image on 2013 January 28 at 11:26 UT is about 70 Mm (Fig.
4d). This value is 27\% less than our estimation of the spine
height from Fig. 4a. Figs 4b-c indicate that the filament is
stable up to the height of 135 Mm.

\subsection[]{Filament on 2013 January 31 in the northern hemisphere}

\begin{figure*}
\includegraphics[width=167mm]{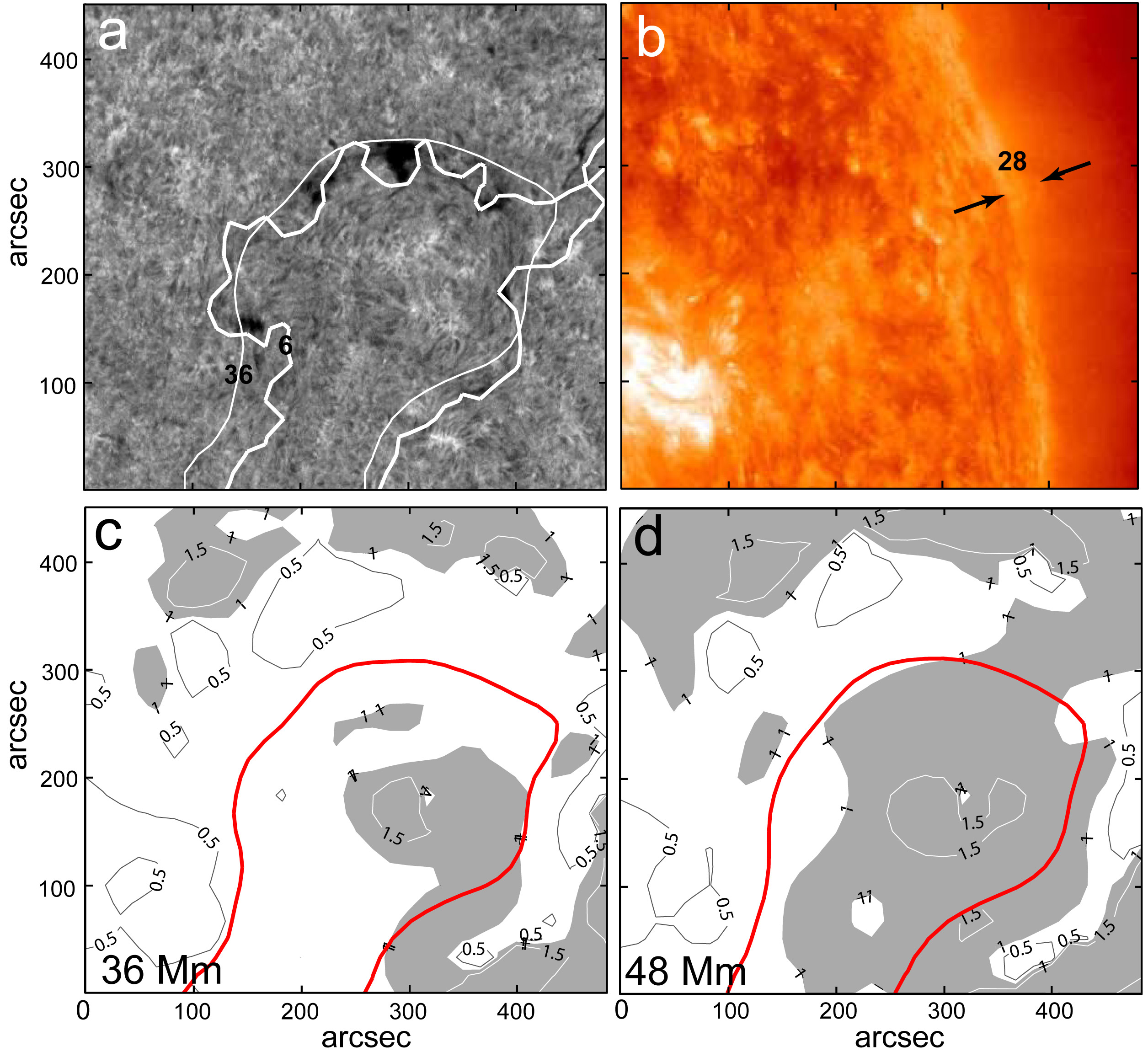}
\caption{H$\alpha$ filtergram of the filament on 2013 January 31
at 20:38 UT with superposed PILs related to the filament at
heights of 6 (thicker line) and 36 Mm (thinner line). (Courtesy of
the Big Bear Solar Observatory). (b) The same filament observed
with {\it STEREO B}/SECCHI EUVI 304-\AA\ on the limb on 2013
January 27  at 21:06 UT. (Courtesy of the {\it STEREO}/SECCHI
Consortium.) (c) - (d) Distributions of the decay index and PILs
(thick red lines) at different heights.}
\end{figure*}

In contrast to the southern filament, the filament in the northern
hemisphere is low one. Its spine fits better to the PIL at a
height of only 36 Mm (Fig. 5a). The height of a faint prominence
observed on limb by the {\it STEREO B} on 2013 January 27 at 21:06
UT is 28 Mm (Fig. 5b). This value is again 22\% less than our
estimation of the spine height obtained from the superposition of
the PILs and filtergram. As it follows from Fig. 5d, instability
may also happen at a rather low height of 48 Mm.

\section[]{Neutral Surface in a Presence of a Flux Rope }

\begin{figure}
\includegraphics[width=84mm]{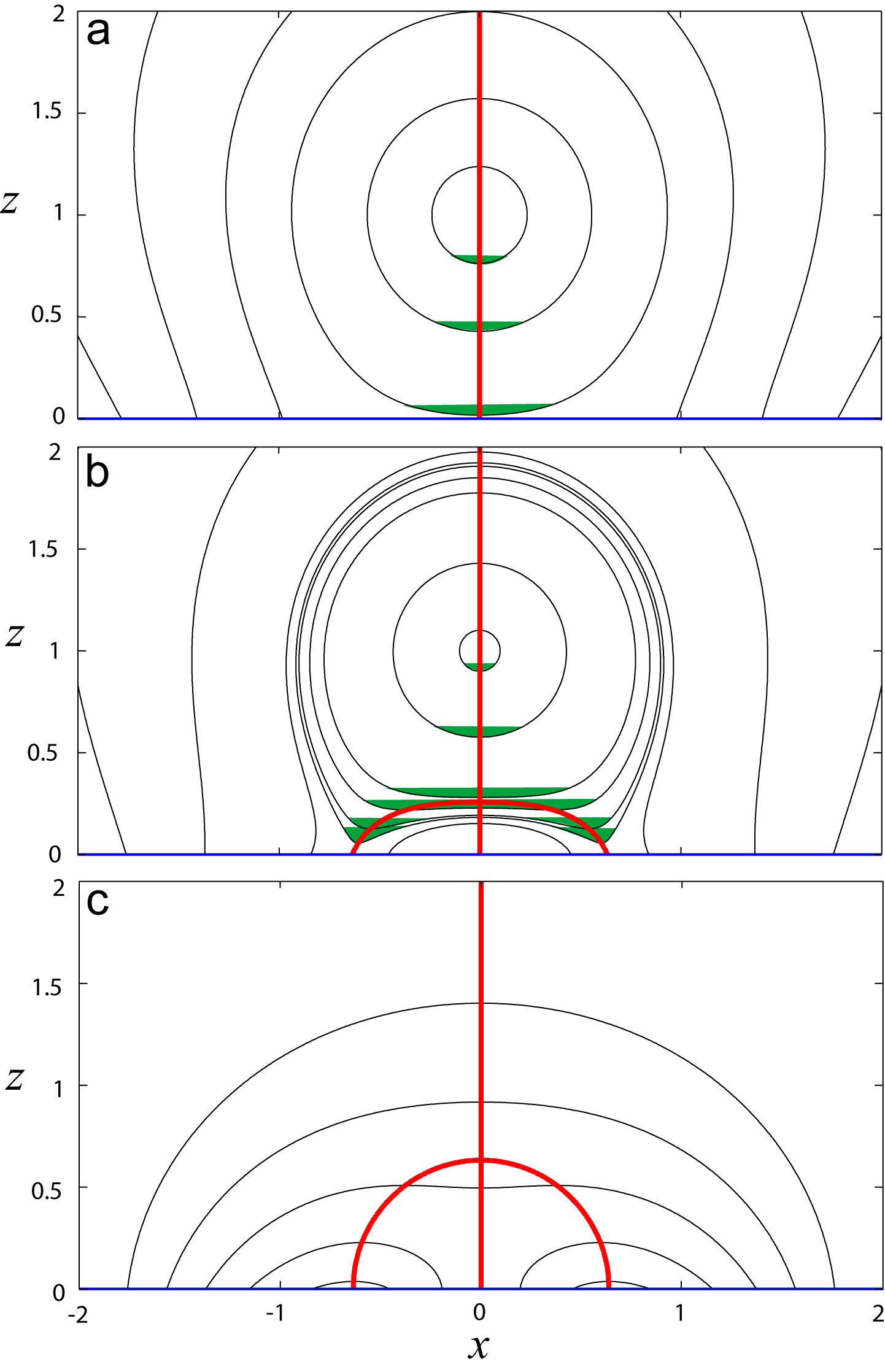}
\caption{ Field lines of a coronal current in equilibrium in a
bipolar field (a), in a bipolar field with a presence of parasitic
polarities (b), field lines of the potential part of the field
shown in (b). Thick (red) lines represent neutral lines $B_z = 0$.
Shadowed (green) areas show possible locations of dense plasma in
dips.}
\end{figure}

Examples presented in the previous section show that the filament
material, as it seem in projection on the disk, fills the space
between two PILs, one at a lower height and the other at a height
of the filament spine. Therefore, the filament material is
distributed near the coronal neutral surface. It seems natural
because a part of the neutral surface where field-line curvature
is directed upward is the location of dips. However, we calculate
only potential part of the coronal field, which is significantly
different from the real coronal field near filaments. Why does the
potential-field neutral surface control the distribution of the
filament plasma?

Let us consider a simple 2D model of electric current equilibrium
in the coronal magnetic field. It was developed basing on the idea
of \citet{b53}  by \citet{b38} for sub-photospheric magnetic
sources represented by a vertical dipole and by \citet{b42}  for
sources represented by a horizontal dipole. If the line electric
current $I$ is directed along the horizontal axis $y$ and the
$z$-axis is vertical, it generates a magnetic field in the upper
hemisphere above a rigid conducting surface

\begin{equation}
B_x^I =  \frac{2I}{c}\left( \frac{z-h}{x^2+(z-h)^2}-
\frac{z+h}{x^2+(z+h)^2} \right),
\end{equation}

 \begin{equation}
B_z^I =  - \frac{2I x}{c}\left( \frac{1}{x^2+(z-h)^2}-
\frac{1}{x^2+(z+h)^2} \right),
\end{equation}
where (0, $h$) are coordinates of the current

Possibly more relevant to typical solar conditions is the coronal
magnetic field created by two sub-photospheric 'charges' $\pm q$
(Filippov, 2013b)

 \begin{equation}
B_x^q =  q \left( \frac{x+a}{(x+a)^2+(z+d)^2}-
\frac{x-a}{(x-a)^2+(z+d)^2} \right),
\end{equation}

 \begin{equation}
B_z^q =  q \left( \frac{z+d}{(x+a)^2+(z+d)^2}-
\frac{z+d}{(x-a)^2+(z+d)^2} \right),
\end{equation}
where $\pm a$ are the $x$-coordinates of the 'charges', $d$ is the
depth below the surface of the photosphere $z = 0$. The electric
current is in equilibrium if

\begin{equation}
I =  \frac{2qahc}{a^2+(h+d)^2}.
\end{equation}

Fig. 6a shows the shape of field lines for $q = 3$, $a = 1.5$, $d
= 1$, $h = 1$ in dimensionless units. Dips exist at bottoms of
nearly circular field lines below the current location. Their
locus is the vertical line $x = 0$, where $B_z = 0$. All filament
material should be close to the symmetry plane. Obviously, that
the neutral surface (line in this 2D case) of potential part of
the field, a symmetric arcade, is the same plane of symmetry $x =
0$.

Then we add a pair of smaller 'charges' $q_1 = \pm 1$ located at
$x = \pm .3$, $z = -.3$ as 'parasitic' polarities. The new field
line pattern shown in Fig. 6b demonstrates the appearance of two
lateral loci of dips. They can be associated with filament barbs
\citep{b2, b10}. The new neutral line is anchored in photospheric
PILs between major and minor polarities on both sides of the
filament and rises into the corona as a flatten arch.

Fig. 6c shows field lines and neutral lines of the potential part
of the configuration with minor polarities. The shape of neutral
lines in Figs 6b and c is nearly the same, while field lines are,
of course, quite different. The arch-like neutral line is anchored
at the same points, although its apex is higher than the apex of
the arch in Fig. 6b. Obviously, the field-line curvature is
directed downward everywhere near the arched potential neutral
line in contrast to Fig. 6b, where the neutral line is the locus
of dips. Nevertheless, we believe that the similarity of the
neutral surfaces in potential and non-potential fields with the
same sub-photospheric sources is the reason for the found tendency
for the filament material to gather near the potential-field
neutral surface. While the structure of the non-potential field in
the corona is quite different from the structure of the potential
field, it is sub-photospheric sources of the coronal magnetic
field that create necessary conditions for plasma accumulation in
certain places.

\section[]{Discussion and Conclusions}
We found that the filament material is distributed in the vicinity
of the coronal potential-field neutral surface. In projection on
the chromosphere, a filament is enclosed between two PILs, one at
a lower height close to the chromosphere and one at a higher
level, which can be considered as a height of the filament spine.
Observations of the same filament on the limb by the {\it STEREO}
spacecraft confirm that the height of the spine is really very
close to the value obtained from the PIL and filament border
matching. Such matching can be used for filament height
estimations in the case of absence of limb observations. First
attempts of filament height estimations on the disk using neutral
surface calculations were made by \citet{b58, b23}, and
\citet{b17}.

Of course, the spine height is not constant along the whole length
of the filament. We choose the PIL that matches the observed
filament spine position over the most part of the length. The
height of this PIL can be considered as the most probable or
averaged height of the spine. The lower side of the filament is
outlined by a low-lying PIL. Filament barbs are housed within
protruding sections of the PIL. However not all PIL protrusions
are filled with filament barbs. These PIL protrusions show
magnetic peninsulas of parasitic polarity on the side of the
filament with the dominating polarity. Ends of filament barbs
touch the low-lying PIL in accordance with results of \citet{b54,
b55}, \citet{b10}, and \citet{b32}.

The main purpose of this paper is to show that the filament
material is located close to the coronal magnetic neutral surface.
According to this purpose we selected rather wide filaments with
developed barbs located close to the central meridian.
Unfortunately, the {\it STEREO} spacecrafts were not in the best
positions for observations of these filaments at the limb. It
needed from 2 to 4 days for the filaments to appear at the limb in
the field of view of {\it STEREO}. During this period of time the
filaments could slightly change their shapes and heights. This is
possibly the major reason of the mismatch between height
estimations by the PIL-method and the {\it STEREO} observations,
which reaches about 20\% in some selected examples. Study of
prominence heights before eruptions \citep{b48a, b26a, b14, b33a}
showed that the height changes slowly for several days within 20\%
mostly due to changes of the prominence shape (appearance and
disappearance of particular threads constituting the prominence).
Prominences start to rise slowly with the speed of few km s$^{-1}$
or less for several hours before the eruption and then are
accelerated fast after reaching the critical height. Filaments and
prominences have not exactly the same boundaries in images taken
in different wavelengths. \citet{b23a}, \citet{b46a}, \citet{b53a}
found that some filaments are wider in EUV than in H$\alpha$. In
coronal EUV lines, prominences are observed in absorption and have
nearly the same appearance as emitting H$\alpha$ prominences.
Sometimes they suddenly change absorption to emission during
activation and eruption \citep{b19a}. In the Transition Region
HeII 304-\AA \ line, prominences look like in H$\alpha$
coronagraphic images \citep{b43a}.

However, there is a more principal source of the mismatch between
the data obtained from on-disc and limb observations. Since a
filament is not a solid object, different parts of it have
different optical thickness in different projections and in
different wavelengths. So we compare the heights of not exactly
the same spatial elements. Nevertheless, filaments, of course, are
quite distinct physical objects although with different visibility
in different observations.

All this variations are most probably within the 20\%
uncertainties that we ascribe to the proposed method of filament
height estimation. This accuracy seems enough for the method to be
useful for the estimation of filament heights on the disc when
there are no observations from other viewpoints. In principle, the
accuracy may be higher but it needs more careful study of
filaments with better conditions of observations (simultaneous
data from two or more viewpoints).

Thus, the filament shape is determined in a first approximation by
the shape of the coronal potential-field neutral surface.
Obviously, that the flux-rope axis, which is believed to coincide
with the filament spine, should be located on the neutral surface.
We may not know the detailed structure of the flux rope containing
the filament but we can judge where filament barbs are expected to
protrude from the filament body, how long and wide could they be.
Condition $B_z = 0$ is the necessary condition for the dip
existence. At low heights where barb ends are observed, the
influence of coronal currents on the $B_z$ distribution is
vanishing because of the neighborhood of the rigid photosphere.
That is why the shape and position of the potential and
non-potential neutral surfaces are very similar (Fig. 6). For this
reason the potential magnetic field calculation is a simple and
valuable tool for analysis of filament morphological properties.
On the other hand, the potential-field decay index distribution
can be used for estimation of the filament stability. As any
coronal magnetic field extrapolation on the base of only one
measured component in the photosphere, our calculations have
limited reliability. More accurate calculations on the base of
vector magnetic field measurements could confirm or disprove our
findings.

\section*{Acknowledgements}

The author thanks the Big Bear Solar Observatory, {\it STEREO},
{\it SDO} teams for the high-quality data supplied. This work was
supported in part by the Russian Foundation for Basic Research
(grant 14-02-92690).

\bsp

\label{lastpage}


\begin{thebibliography}{99}


\bibitem[\protect\citeauthoryear{Antiochos, Dahlburg \& Klimchuk}{Antiochos et al.}{1994}]{b1}
Antiochos S. K., Dahlburg R. B., Klimchuk J. A., 1994, ApJ, 420,
L41
\bibitem[\protect\citeauthoryear{Aulanier \& D$\acute{e}$moulin} {1998}]{b2}
Aulanier G., D$\acute{e}$moulin P., 1998, A\&A, 329, 1125
\bibitem[\protect\citeauthoryear{Aulanier et al.} {1998}]{b3}
Aulanier G., D$\acute{e}$moulin P., van Driel-Gesztelyi L., Mein
P., Deforest C., 1998, A\&A, 335, 309
\bibitem[\protect\citeauthoryear{Aulanier et al.} {1999}]{b4}
Aulanier G., D$\acute{e}$moulin P., Mein N., van Driel-Gesztelyi
L., Mein P., Schmieder B., 1999, A\&A,  342, 867.
\bibitem[\protect\citeauthoryear{Aulanier, DeVore \& Antiochos}{Aulanier et al.} {2006}]{b5}
Aulanier G., DeVore C. R., Antiochos S. K., 2006, ApJ, 646, 1349
\bibitem[\protect\citeauthoryear{Babcock \& Babcock}{1955}]{b6}
Babcock H. W., Babcock H. D., 1955, ApJ, 121, 349
\bibitem[\protect\citeauthoryear{Bommier et al.}{1994}]{b7}
Bommier V., Landi Degl'Innocenti E., Leroy J.-L., Sahal-Brechot
S., 1994, Sol. Phys., 154, 231
\bibitem[\protect\citeauthoryear{Bommier \& Leroy}{1998}]{b8}
Bommier V., Leroy J. L., 1998, in  Webb D.,  Rust D., Schmieder
B., eds, ASP Conf. Ser. Vol. 150, IAU Colloq. 167: New
Perspectives on Solar Prominences, Astron. Soc. Pac., San
Francisco, p. 434
\bibitem[\protect\citeauthoryear{Chae et al.}{2001}]{b9}
Chae J., Wang H., Qiu J., Goode P. R., Strous L.,
Yun H. S., 2001, ApJ,  560, 476
\bibitem[\protect\citeauthoryear{Chae, Moon \& Park }{Chae et al.}{2005}]{b10}
Chae J., Moon Y.-J., Park Y.-D., 2005, ApJ, 626, 574
\bibitem[\protect\citeauthoryear{d'Azambuja \& d'Azambuja}{1948}]{b11}
d'Azambuja M., d'Azambuja L., 1948, Ann. Observ. Paris, Meudon 6,
Fasc. VII
\bibitem[\protect\citeauthoryear{DeVore \& Antiochos}{2000}]{b12}
DeVore C. R., Antiochos S. K., 2000, ApJ, 539, 954
\bibitem[\protect\citeauthoryear{Filippov}{2011}]{b13}
Filippov B. P., 2011, Astron. Rep., 55, 541
\bibitem[\protect\citeauthoryear{Filippov}{2013a}]{b14}
Filippov B., 2013a, ApJ, 773, 10
\bibitem[\protect\citeauthoryear{Filippov}{2013b}]{b15}
Filippov B., 2013b, Sol. Phys., 283, 401
\bibitem[\protect\citeauthoryear{Filippov}{2015a}]{b16}
Filippov B., 2015a, Astron. Rep. (in press)
\bibitem[\protect\citeauthoryear{Filippov}{2013b}]{b17}
Filippov B., 2015b, Geomagnetism \& Aeronomy (in press)
\bibitem[\protect\citeauthoryear{Filippov \& Den}{2000}]{b18}
Filippov B. P., Den O. G., 2000,  Astron. Lett., 26, 322
\bibitem[\protect\citeauthoryear{Filippov \& Den}{2001}]{b19}
Filippov B. P., Den O. G., 2001, J. Geophys. Res., 106, 25177
\bibitem[\protect\citeauthoryear{Filippov \& Koutchmy}{2002}]{b19a}
Filippov B., Koutcmy S., 2002, Sol. Phys., 208, 283
\bibitem[\protect\citeauthoryear{Filippov, Gopalswamy \& Lozhechkin}{Filippov et al.}{2001}]{b20}
Filippov B. P, Gopalswamy N., Lozhechkin A. V., 2001, Sol. Phys.,
203, 119
\bibitem[\protect\citeauthoryear{Filippov, Gopalswamy \& Lozhechkin}{Filippov et al.}{2002}]{b21}
Filippov B. P., Gopalswamy N., Lozhechkin A. V., 2002, Astron.
Rep., 46, 417
\bibitem[\protect\citeauthoryear{Gibson \& Fan}{2006}]{b22}
Gibson S. E., Fan Y., 2006, J. Geophys. Res., 111, A12103
\bibitem[\protect\citeauthoryear{Grechnev et al.}{2014}]{b23}
Grechnev V. V., Uralov A. M., Slemzin V. A., Chertok I. M.,
Filippov B. P., Rudenko G. V., Temmer, M., 2014, Sol. Phys., 289,
289
\bibitem[\protect\citeauthoryear{Heinzel, Schmieder \& Tziotziou}{Heinzel et al.}{2001}]{b23a}
Heinzel P., Schmieder B., Tziotziou K., 2001, ApJL, 561, L223
\bibitem[\protect\citeauthoryear{Howard \& Harvey}{1964}]{b24}
Howard R. F., Harvey J. W., 1964, ApJ, 139, 1328
\bibitem[\protect\citeauthoryear{Howard et al.}{2008}]{b25}
Howard R. A. et al., 2008, Space Sci. Rev., 136, 67
\bibitem[\protect\citeauthoryear{Ioshpa}{1968}]{b26}
Ioshpa B. A., 1968, in  Kipenheuer K.O., ed.,  IAU Symp. 35:
Structure and Development of Solar Active Region, D. Reidel,
Dordrecht, p. 261
\bibitem[\protect\citeauthoryear{Joshi \& Srivastava}{2011}]{b26a}
Joshi A. D., Srivastava N., 2011, ApJ, 730, 104
\bibitem[\protect\citeauthoryear{Kim}{1990}]{b27}
Kim I. S., 1990, in Ruzdjak V., Tandberg-Hanssen E., eds, Dynamics
of Quiescent Prominences, Lecture Notes in Physics, Vol. 363,
Springer, Berlin, p. 49
\bibitem[\protect\citeauthoryear{Kippenhahn \& Schl\"{u}ter}{1957}]{b28}
Kippenhahn R., Schl\"{u}ter A., 1957, Z. Astrophys., 43, 36
\bibitem[\protect\citeauthoryear{Kuperus \& Raadu}{1974}]{b29}
Kuperus M., Raadu M. A., 1974,  A\&A, 31, 189
\bibitem[\protect\citeauthoryear{Leroy}{1977}]{b30}
Leroy J. L., 1977, A\&A, 60, 79
\bibitem[\protect\citeauthoryear{Leroy}{1978}]{b31}
Leroy J. L., 1978, A\&A, 64, 247
\bibitem[\protect\citeauthoryear{Lin et al.}{2005}]{b32}
Lin Y., Wiik J. E., Engvold O., Rouppe van der Voort L., Frank Z.
A., 2005, Sol. Phys., 227, 283
\bibitem[\protect\citeauthoryear{Mackay et al.}{2010}]{b33}
Mackay D. H., Karpen J. T., Ballester J. L., Schmieder B.,
Aulanier G., 2010, Space Sci. Rev., 151, 333
\bibitem[\protect\citeauthoryear{McCauley et al.}{2015}]{b33a}
McCauley P. I., Su Y. N., Schanche N., Evans K. E., Su C.,
McKillop S., Reeves K. K., 2015, Sol. Phys., 290, 1703
\bibitem[\protect\citeauthoryear{Martin}{1998}]{b34}
Martin S. F., 1998, Sol. Phys., 182, 107
\bibitem[\protect\citeauthoryear{Martin \& Echols}{1994}]{b35}
Martin S. F., Echols C. R., 1994, in Rutten R. J.,  Schrijver C.
J., eds, Solar Surface Magnetism, Kluwer, Dordrecht, p. 339
\bibitem[\protect\citeauthoryear{Martin, Bilimoria \& Tracadas}{Martin et al.}{1994}]{b36}
Martin S. F., Bilimoria R., Tracadas P. W., 1994, in  Rutten R.
J., Schrijver C. J., eds, Solar Surface Magnetism,  Kluwer,
Dordrecht, p. 303
\bibitem[\protect\citeauthoryear{McIntosh}{1972}]{b37}
McIntosh P. S., 1972, Rev. Geophys. Space Phys., 10, 837
\bibitem[\protect\citeauthoryear{Molodenskii \& Filippov}{1987}]{b38}
Molodenskii M. M., Filippov B. P., 1987, Soviet Astron., 31, 564
\bibitem[\protect\citeauthoryear{Parenti}{2014}]{b39}
Parenti S., 2014, Living Rev. Solar Phys., 11, 1
\bibitem[\protect\citeauthoryear{Plocieniak \& Rompolt}{1973}]{b40}
Plocieniak S., Rompolt B., 1973, Sol. Phys., 29, 399
\bibitem[\protect\citeauthoryear{Pneuman}{1983}]{b41}
Pneuman G. W., 1983, Sol. Phys., 88, 219
\bibitem[\protect\citeauthoryear{Priest \& Forbes}{1990}]{b42}
Priest E. R., Forbes T. G., 1990, Sol. Phys., 126, 319
\bibitem[\protect\citeauthoryear{Priest, Hood \& Anzer}{Priest et al.}{1989}]{b43}
Priest E. R., Hood A. W., Anzer U., 1989,  ApJ,  344, 1010
\bibitem[\protect\citeauthoryear{Romeuf et al.}{2007}]{b43a}
Romeuf D., Meunier N., No$\ddot{e}$ns J.-C., Koutchmy S., Jimenez
R., Wurmser O., Rochain S.,'Observateurs Associ$\acute{e}$s' Team,
2007, AA, 462, 731
\bibitem[\protect\citeauthoryear{Rompolt}{1990}]{b44}
Rompolt B., 1990, Bull. Hvar Obs. 14, 37
\bibitem[\protect\citeauthoryear{Rust}{1967}]{b45}
Rust D. M., 1967, ApJ, 150, 313
\bibitem[\protect\citeauthoryear{Rust \& Kumar}{1994}]{b46}
Rust D. M., Kumar A., 1994, Sol. Phys., 155, 69
\bibitem[\protect\citeauthoryear{Schmieder et al.}{2004}]{b46a}
Schmieder B., Lin Y., Heinzel P., Schwartz P., 2004, Sol. Phys.,
221, 297
\bibitem[\protect\citeauthoryear{Schou et al.}{2012}]{b47}
Schou, J. et al., 2012, Sol. Phys., 275, 229
\bibitem[\protect\citeauthoryear{Smith \& Ramsey}{1967}]{b48}
Smith S. F., Ramsey H. E., 1967, Sol. Phys., 2, 158
\bibitem[\protect\citeauthoryear{Sterling, Moore \& Freeland}{Sterling et al.}{2011}]{b48a}
Sterling A. C., Moore R. L., Freeland S. L., 2011, ApJL, 731, L3
\bibitem[\protect\citeauthoryear{Tandberg-Hanssen}{1970}]{b49}
Tandberg-Hanssen E., 1970, Sol. Phys., 15, 359
\bibitem[\protect\citeauthoryear{Tandberg-Hanssen}{1974}]{b50}
Tandberg-Hanssen E., 1974, Solar Prominences, Geophysics and
Astrophysics Monographs 12, D. Reidel, Dordrecht.
\bibitem[\protect\citeauthoryear{van Ballegooijen}{2004}]{b51}
van Ballegooijen A. A., 2004, ApJ,  612, 519
\bibitem[\protect\citeauthoryear{van Ballegooijen \& Martens}{1989}]{b52}
van Ballegooijen A. A.,  Martens, P. C. H., 1989, ApJ, 343, 971
\bibitem[\protect\citeauthoryear{van Tend \& Kuperus}{1978}]{b53}
van Tend W., Kuperus M., 1978, Sol. Phys., 59, 115
\bibitem[\protect\citeauthoryear{Vial et al.}{2012}]{b53a}
Vial J.-C., Olivier K., Philippon A. A., Vourlidas A., Yurchyshyn
V., 2012, AA, 541, A108
\bibitem[\protect\citeauthoryear{Wang}{1999}]{b54}
Wang Y.-M., 1999, ApJ,  520, L71
\bibitem[\protect\citeauthoryear{Wang}{2001}]{b55}
Wang Y.-M., 2001, ApJ, 560, 456
\bibitem[\protect\citeauthoryear{Wuelser et al.}{2004}]{b56}
Wuelser J.-P. et al., 2004, in Fineschi S, Gummin M. A., eds,
Proc. SPIE Conf. Ser. Vol. 5171, Telescopes and Instrumentation
for Solar Astrophysics. SPIE, Bellingham, p. 111
\bibitem[\protect\citeauthoryear{Zirin \& Severny}{1961}]{b57}
Zirin H., Severny A., 1961, The Observatory, 81, 155
\bibitem[\protect\citeauthoryear{Zagnetko, Filippov \& Den}{Zagnetko et al.}{2005}]{b58}
Zagnetko A. M., Filippov B. P., Den O. G., 2005, Astron. Rep. 49,
425



\end{thebibliography}
\end{document}